\documentclass[apj]{emulateapj}
\usepackage{graphicx}
\usepackage{natbib}
\usepackage{color}

\shorttitle{SN~2016gkg}
\shortauthors{Tartaglia et al.}

\newcommand{\h}{\ion{H}{1}}
\newcommand{\ha}{H$\alpha$}
\newcommand{\he}{\ion{He}{1}}

\newcommand{\kms}{\,km s$^{-1}$}

\newcommand{\msun}{\,M$_{\sun}$}
\newcommand{\rsun}{\,R$_{\sun}$}

\begin{document} 

\title{The progenitor and early evolution of the Type IIb SN~2016gkg}

\author{
L. Tartaglia\altaffilmark{1,2}, 
M. Fraser\altaffilmark{3}, 
D.J. Sand\altaffilmark{1}, 
S. Valenti\altaffilmark{2}, 
S. J. Smartt\altaffilmark{4}, 
C. McCully\altaffilmark{6,7},
J. P. Anderson\altaffilmark{5},
I. Arcavi\altaffilmark{6,7,18},
N. Elias-Rosa\altaffilmark{8},
L. Galbany\altaffilmark{9},
A. Gal-Yam\altaffilmark{10},
J.B. Haislip\altaffilmark{11}, 
G. Hosseinzadeh\altaffilmark{6,7}, 
D. A. Howell\altaffilmark{6,7},
C. Inserra\altaffilmark{4},
S. W. Jha\altaffilmark{12}, 
E. Kankare\altaffilmark{4},
P. Lundqvist\altaffilmark{13}, 
K. Maguire\altaffilmark{14,4}, 
S. Mattila\altaffilmark{15}, 
D. Reichart\altaffilmark{11},
K. W. Smith\altaffilmark{4},
M. Smith\altaffilmark{16}, 
M. Stritzinger\altaffilmark{17}, 
M. Sullivan\altaffilmark{16}, 
F. Taddia\altaffilmark{13}, 
L. Tomasella\altaffilmark{8}
}

%
\begin{abstract}
We report initial observations and analysis on the Type IIb SN~2016gkg in the nearby galaxy NGC~613. SN~2016gkg exhibited a clear double-peaked light curve during its early evolution, as evidenced by our intensive photometric follow-up campaign.
SN~2016gkg shows strong similarities with other Type IIb SNe, in particular with respect to the \he~emission features observed in both the optical and near infrared.  
SN~2016gkg evolved faster than the prototypical Type~IIb SN~1993J, with a decline similar to that of SN~2011dh after the first peak.
The analysis of archival {\it Hubble Space Telescope} images indicate a pre-explosion source at SN~2016gkg's position, suggesting a progenitor star with a $\sim$mid F spectral type and initial mass $15-20$\msun, depending on the distance modulus adopted for NGC~613.
Modeling the temperature evolution within $5\,\rm{days}$ of explosion, we obtain a progenitor radius of $\sim\,48-124$\rsun, smaller than that obtained from the analysis of the pre-explosion images ($240-320$\rsun).
\end{abstract}


\altaffiltext{1}{Texas Tech University, Physics Department, Box 41051, Lubbock, TX 79409-1051, USA; ltartaglia@ucdavis.edu}
\altaffiltext{2}{Department of Physics, University of California, Davis, CA 95616, USA}
\altaffiltext{3}{School of Physics, O'Brien Centre for Science North, University College Dublin, Belfield, Dublin 4, Ireland.}
\altaffiltext{4}{Astrophysics Research Centre, School of Mathematics and Physics, Queens University Belfast, Belfast BT7 1NN, UK}
\altaffiltext{5}{European Southern Observatory, Alonso de C\'ordova 3107, Casilla 19, Santiago, Chile}
\altaffiltext{6}{Department of Physics, University of California, Santa Barbara, CA ,93106-9530, USA}
\altaffiltext{7}{Las Cumbres Observatory Global Telescope Network, 6740 Cortona Dr., Suite 102, Goleta, CA 93117, USA}
\altaffiltext{8}{INAF Osservatorio Astronomico di Padova, Vicolo dell’Osservatorio 5, 35122 Padova, Italy}
\altaffiltext{9}{PITT PACC, Department of Physics and Astronomy, University of Pittsburgh, Pittsburgh, PA 15260, USA}
\altaffiltext{10}{Benoziyo Center for Astrophysics, Faculty of Physics, Weizmann Institute of Science, Rehovot 76100, Israel}
\altaffiltext{11}{University of North Carolina at Chapel Hill, Campus Box 3255, Chapel Hill, NC 27599-3255, USA}
\altaffiltext{12}{Department of Physics and Astronomy, Rutgers, the State University of New Jersey, 136 Frelinghuysen Road, Piscataway, NJ 08854, USA}
\altaffiltext{13}{Department of Astronomy and The Oskar Klein Centre, AlbaNove University Center, Stockholm University, SE-106 91 Stockholm, Sweden}
\altaffiltext{14}{European Organisation for Astronomical Research in the Southern Hemisphere (ESO), Karl-Schwarzschild-Str. 2, D-85748 Garching b. M\"unchen, Germany}
\altaffiltext{15}{Tuorla Observatory, Department of Physics and Astronomy, University of Turku, V\"ais\"al\"antie 20, FI-21500 Piikki\"o, Finland}
\altaffiltext{16}{School of Physics and Astronomy, University of Southampton, Southampton, SO17 1BJ, UK}
\altaffiltext{17}{Department of Physics and Astronomy, Aarhus University, Ny Munkegade 120, 8000 Aarhus C, Denmark}
\altaffiltext{18}{Einstein Fellow}

\section{Introduction}
\label{sec:intro}
Type IIb supernovae (SNe IIb) are likely the result of the core collapse of massive stars which have lost most of their hydrogen envelope prior to explosion.  
At early times, and through maximum light, SN IIb show hydrogen features typical of Type II SNe, which later give way to He I absorption lines similar to those observed in SNe Ib \citep{1994AJ....108.2220F}. 

Several SNe IIb caught soon after explosion exhibited `double-peaked' light curves.
These have been interpreted as the signature of shock breakout cooling of a progenitor star core surrounded by extended, low-mass material \citep[e.g.][]{2012ApJ...757...31B,Nakar14}. 
Well-studied examples of this phenomenon include SNe 1993J \citep[e.g.][]{Richmond94}, 2011dh \citep[e.g.][]{Arcavi11,Ergon14}, 2011fu \citep{2013MNRAS.431..308K}  and 2013df \citep{morales14,Vandyk14}. 
Progenitor constraints from deep pre- and post-explosion imaging \citep[e.g.][for SN~1993J]{Aldering94,Maund09} have revealed a picture of Type IIb's originating from yellow massive stars ($M_{ZAMS}=12-16$\msun) in binary systems. 
\begin{figure*}
\begin{center}
\mbox{ \epsfysize=10.0cm \epsfbox{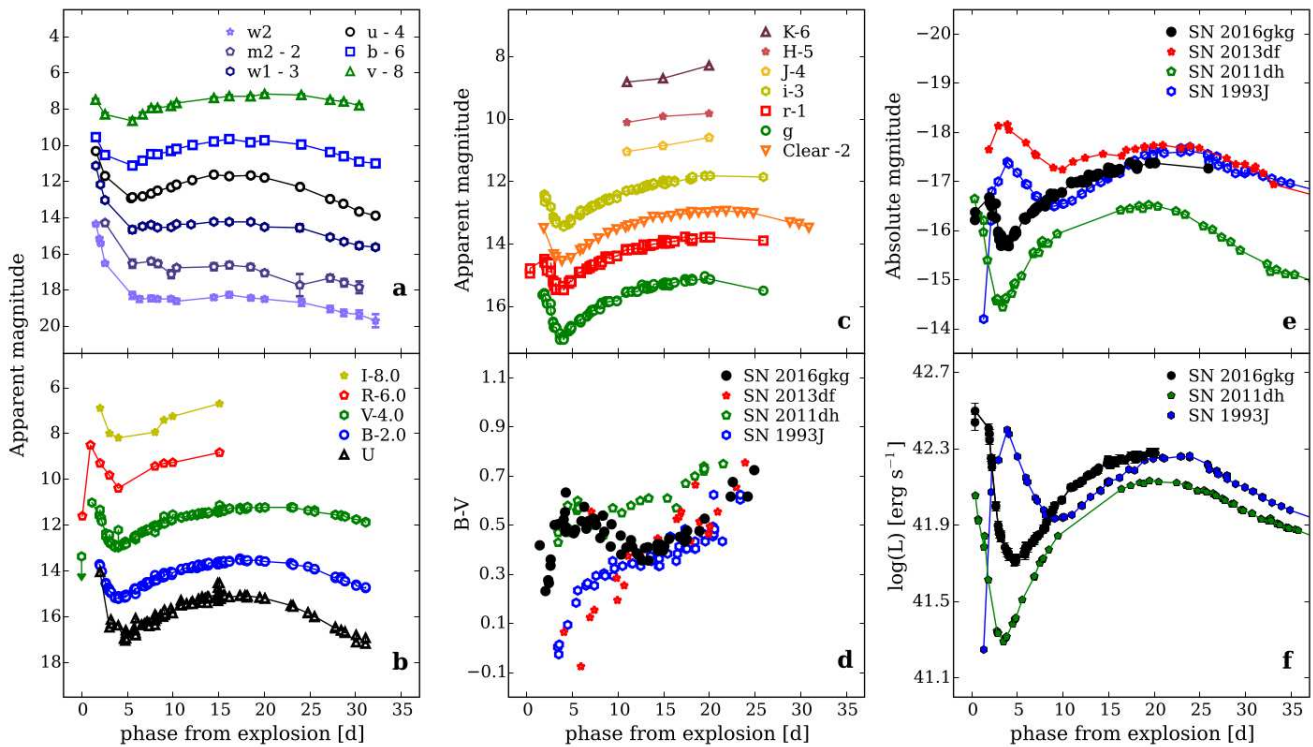}} 
\caption{{\bf (a)} UVOT, {\bf (b)} $UBVRI$ and {\bf (c)} and $grizJHK$ light-curves of SN~2016gkg. 
{\bf (d)} $B-V$ color evolution compared with those of SNe~1993J, 2011dh and 2013df. {\bf (e)} Absolute $r$ light curves compared with those of SNe~1993J ($R$-band), 2011dh ($g$-band) and 2013df ($R$-band).
{\bf (f)} {\bf Pseudo-bolometric} light curve compared to those of SNe~2011dh and 1993J. 
The Pseudo-bolometric light curves were computed using Simpson's rule, integrating the photometric data.
At early phase only $Vr$ bands are available. For the missing bands, we assumed constant colors and a 10\% uncertainty on the computed flux. Errors for SN~2016gkg were computed assuming a fixed distance modulus ($\mu=32.11\,\rm{mag}$).
The data in panels a,b,c,f are available as Data behind the Figure. \label{fig:LC}}
\end{center}
\end{figure*}

Here we present the first month of evolution of the Type IIb SN~2016gkg, progenitor constraints from pre-explosion {\it HST} imaging and its early temperature evolution.
SN~2016gkg was discovered by V. Buso on 2016 September 20.19 UT\footnote{\url{http://ooruri.kusastro.kyoto-u.ac.jp/mailarchive/vsnet-alert/20188}}, 78.4$"$ South and 49.1$"$ West from the nucleus of NGC~613. 
It was confirmed via photometry \citep{Nicholls16,Tonry16} and typed as a young Type II SN \citep{Jha16}. 
After a fast decline \citep{Chen16}, its light curve began to rise again towards a second maximum. 
Progenitor constraints from {\it Hubble Space Telescope} ({\it HST}) pre-imaging at the location of SN~2016gkg were also reported \citep{Kilpatrick16}, and we will discuss these further below. 
We adopt a distance of $26\,\rm{Mpc}$, based on a Tully-Fisher measurement \citep[$m-M=32.1\pm0.4\,\rm{mag}$;][]{Tully09}. 
However, since Tully-Fisher measurements for NGC~613 range from $\sim20-30\,\rm{Mpc}$, we will also discuss the implications of a lower host distance ($20\,\rm{Mpc}$). 
We assume $A_V=0.053\,\rm{mag}$ for the foreground Galactic extinction \citep{Schlafly11}.

\section{Observations And Data Reduction} \label{sec:obs}
\subsection{Photometry}
Imaging data were processed as follows.
$UBVgri$ frames from the Las Cumbres Observatory global telescope network \citep[LCO;][]{lcogt} were reduced using {\sc lcogtsnpipe} \citep[e.g.][]{Valenti16}. 
$UBVRI$ data from the Public ESO Spectroscopic Survey of Transient Objects (PESSTO\footnote{\url{http://www.pessto.org/}}) using the ESO Faint Object Spectrograph and Camera v2 (EFOSC2) were reduced using the PESSTO pipeline \citep{pessto}. 
$gri$ data were also obtained by the Nordic Opitcal Telescope (NOT) Unbiased Transient Survey (NUTS\footnote{\url{http://csp2.lco.cl/not/}}) with the Andalucia Faint Object Spectrograph and Camera (ALFOSC) on the NOT and reduced using a dedicated pipeline \citep[SNOoPY;][]{snoopy}.
Filterless data from the $D<40\,\rm{Mpc}$ (DLT40) SN search using the PROMPT5 telescope \citep{prompt} were calibrated to APASS r-band\footnote{\url{http://www.aavso.org/apass}}. 
Early {\it Swift} data (PIs Brown, Drout) were reduced following the prescriptions of \citet{Brown09}, using the zero-points from \citet{Breeveld10}. Data from the Asteroid Terrestrial-impact Last Alert System (ATLAS) telescope system \citep{ATLAS} were also used in our early time light curve \citep{2016arXiv161106451A}.

The light curves are presented in Fig.~\ref{fig:LC}, including publicly available early time photometry from ASAS-SN and other sources \citep{Nicholls16}.

\subsection{Spectroscopy}
\begin{figure*}
\begin{center}
\mbox{ \epsfysize=11.0cm \epsfbox{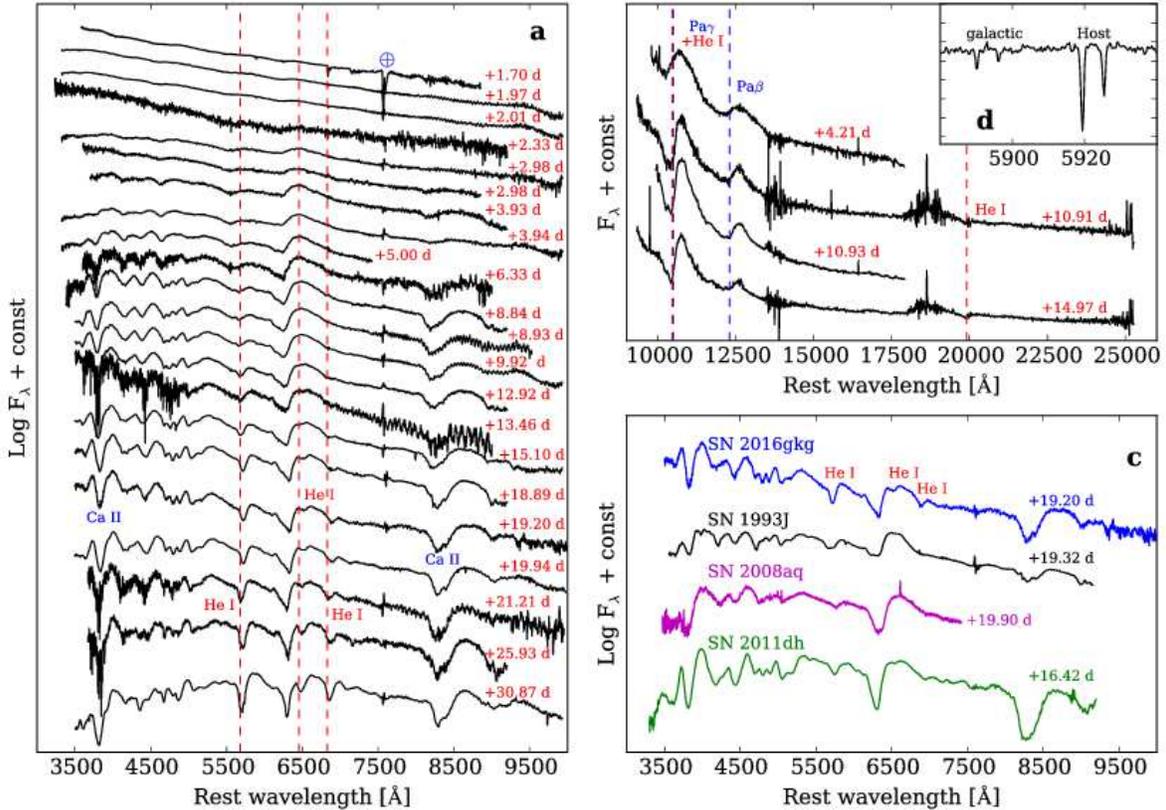}} 
\caption{{\bf (a)} Spectral sequence of SN~2016gkg. Helium absorption lines are marked.
{\bf (b)} NIR spectra. Blue and red dashed lines mark \he~and \h~absorption minima respectively. {\bf (c)} Comparison of SN~2016gkg with other SN~IIb at similar phase: SNe~1993J, 2008aq \citep{2014AJ....147...99M} and 2011dh. 
{\bf (d)} High-resolution \ion{Na}{1D} features.
\label{fig:spec}}
\end{center}
\end{figure*}

Spectra were obtained from multiple sources, reduced in a standard manner.
The classification spectrum \citep{Jha16} was obtained using the Southern African Large Telescope with the Robert Stobie Spectrograph and reduced using \textsc{pysalt} \citep{2010SPIE.7737E..25C}.
Optical spectra from PESSTO were obtained with EFOSC2, while NIR spectra were taken with the Son OF ISAAC camera \citep[SOFI;][]{1998Msngr..91....9M}; all of these data were reduced as in \citet{pessto}. FLOYDS optical spectra from LCOGT were reduced as in \citet{Valenti14}. Optical spectra from ALFOSC and the NUTS collaboration were reduced using their dedicated pipeline (FOSCGUI\footnote{\url{http://sngroup.oapd.inaf.it/foscgui.html}}).
Gemini South data using the NIR spectrograph FLAMINGOS-2 ($JH$ grism, $1.0-1.8\,\mu\rm{m}$ and $R\approx1000$) were reduced using the {\sc F2} {\sc pyraf} package. 
The final sequence is presented in Fig.~\ref{fig:spec}. Spectra will be released through the Weizmann Interactive Supernova data REPository \citep[][]{2012PASP..124..668Y}.

\subsection{{\it HST} pre-explosion imaging}
NGC~613 was observed with the {\it HST} Wide-Field Planetary Camera 2 (WFPC2) on 2001 August 21 (PID: 9042; PI: S.~Smartt). 
Exposures of $2\times160\,\rm{s}$ were taken in the {\it F450W}, {\it F606W} and {\it F814W} filters.
The SN position falls on the WF4 chip, with a pixel scale of 0.1\arcsec/px, where it lies $\sim$\/$10\,\rm{pixels}$ from the CCD edge.

\section{Analysis And Results}
\subsection{Light Curves} \label{sec:phot}
We will adopt the discovery as the explosion epoch: $\rm{JD}=2457651.69028$. This is consistent with the explosion epoch obtained using the \citealt{2016arXiv160703700S} model to reproduce the early phase photometry (see \citealt{2016arXiv161106451A}).

After an initial maximum, the light curves show a fast decline up to $\simeq+4\,\rm{d}$ ($\rm{JD}\simeq2457655$), when the optical magnitudes increase again toward the main peak.
In the {\it Swift} bands (Fig.~\ref{fig:LC}a) we note a similar decline, although the rise toward the second peak is less pronounced, with the $UVW1$, $UVM2$ and $UVW2$ curves flattening after the first peak. 
A similar behavior was observed for SN~2011dh, but not for SN~2013df, where the UV light curves showed an almost linear decline after maximum \citep[][]{2015MNRAS.454...95M}.

The fast early decline of SN~2016gkg is highlighted in Fig.~\ref{fig:LC}e where the absolute light curve is compared to those of other Type IIb SNe \citep[SNe~1993J, 2011dh and 2013df;][]{1995A&AS..110..513B,Arcavi11,2015MNRAS.454...95M}.
The $g-$ and $B-$band decline rates of SN~2016gkg ($0.58\,\rm{mag/d}$ and $0.82\,\rm{mag/d}$ within the first $\sim$\/$2\,\rm{d}$ after first maximum) are greater than those of SN~1993J in $B$ ($0.31 \rm{mag/d}$ within $\sim5\,\rm{d}$), but slightly lower than those observed in SN~2011dh ($0.86\,g-\rm{mag/d}$ within $\sim$\/$3\,\rm{d}$ from the first maximum).

We infer a peak absolute magnitude of $M_B=-16.48\pm0.38\,\rm{mag}$ (where the errors in absolute magnitude here are driven by the NGC~613 distance uncertainty), rapidly decreasing to $M_B=-15.09\pm0.38\,\rm{mag}$ within the first $\sim$\/$2\,\rm{days}$ of evolution. 
After the first peak, the absolute $B$-magnitudes are comparable with those of other SN IIb, with an absolute magnitude for the second peak of $M_B=-17.03\pm0.38\,\rm{mag}$. 

In Fig.~\ref{fig:LC}d we compare the early $B-V$ color evolution with other Type IIb SNe.
While the early evolution is similar to that observed in SN~1993J and SN~2011dh, from $\simeq+5\,\rm{d}$, we note an unusual flattening in the color evolution, which is not observed in other Type IIb SNe. 

In Fig.~\ref{fig:LC}f we show the pseudo-bolometric light curve of SNe~2016gkg, 2011dh and 1993J, obtained by integrating the UV through NIR fluxes. 
While the overall morphology of SNe~2016gkg and 2011dh is similar, the first maximum of SN~2016gkg is significantly brighter, even though their second maxima are consistent. 
The different decline rates implies a diversity in outer envelope extent for the three progenitors (lower in the case of SN~2011dh, larger for SN~1993J).  

\begin{figure*}
\begin{center}
\mbox{ \epsfysize=6.5cm \epsfbox{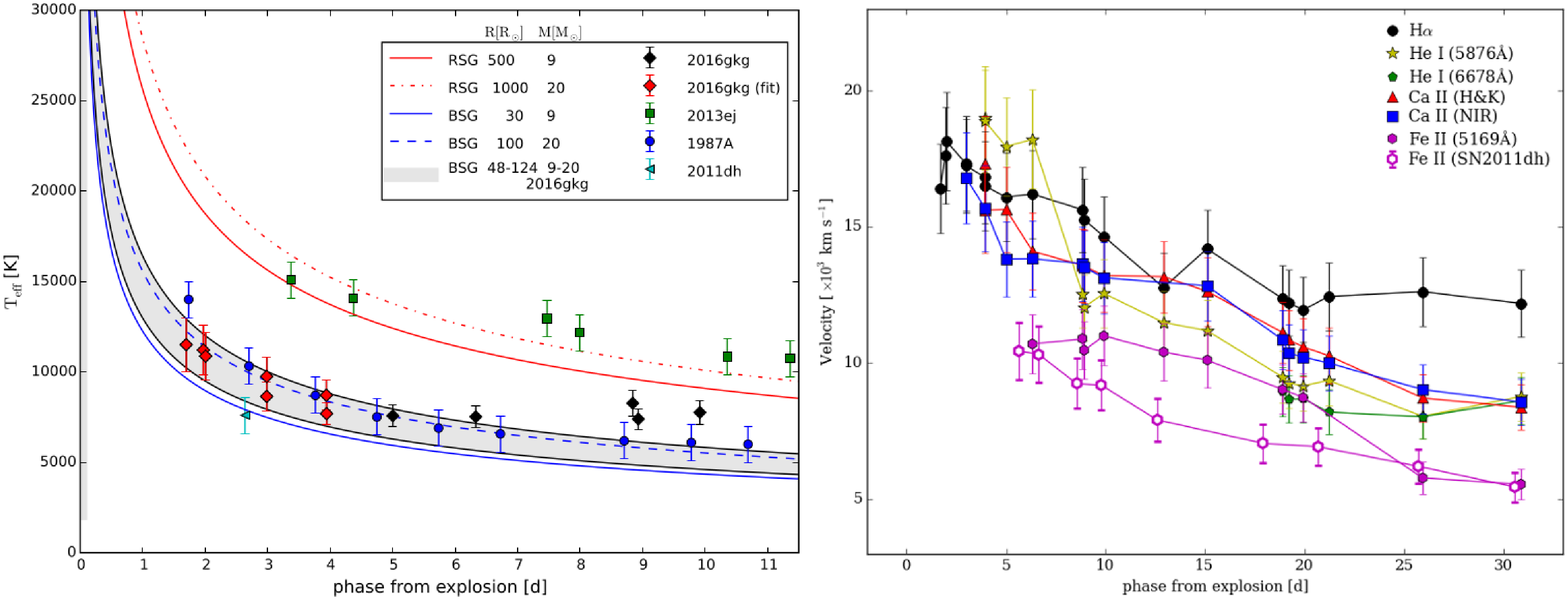}} 
\caption{{\bf (a)} Progenitor radius constraints using the formalism of \cite{Rabinak11} for RSG (red line), BSG (blue line) and 2016gkg (black points) and compared to those of SN~2011dh, SN~1987A \citep{1987MNRAS.227P..39M}, and SN 2013ej \citep{Valenti14}. The fit for SN~2016gkg was limited to $\sim+5\,\rm{d}$ from the explosion (red diamonds), following the prescriptions of \cite{Rabinak11}. {\bf (b)} Expansion velocity evolution for different lines. \ion{Fe}{2} (5169\AA) velocity evolution of SN~2011dh is also shown for comparison. The velocities were computed measuring the positions of the minima of the P-Cygni absorption components.
\label{fig:radius}}
\end{center}
\end{figure*}

\subsection{Spectra} \label{sec:spec}
The spectral fluxes were adjusted using photometry at similar epochs, and corrected for Galactic foreground extinction and redshift of the host \citep[$v_{\sun}=1481\pm5$\kms;][]{Koribalski04}.
At early phases (i.e. $\rm{phase}\lesssim2.7\,\rm{d}$) spectra are dominated by a blue, almost featureless continuum, while P-Cygni line profiles become evident as the temperature of the ejecta decreases. 

Fitting a black-body to the spectral continuum, we infer a temperature of $\simeq10000\,\rm{K}$ at $+1.70\,\rm{d}$, rapidly decreasing to $\simeq7000\,\rm{K}$ at $\sim+5\,\rm{days}$. 
We also estimated the host-galaxy reddening at the position of SN~2016gkg using the equivalent width of the \ion{Na}{1D} doublet in an XSHOOTER spectrum ($\rm{EW}(D1)=0.26\pm0.02\AA$, $\rm{EW}(D2)=0.42\pm0.02\AA$; see Fig.~\ref{fig:spec}d). Based on the correlation with the color excess \citep[e.g.][]{Poznanski12}, we obtain $E(B-V)_{\rm{NGC\,613}}\lesssim0.15\,\rm{mag}$.
Including this contribution, we get a slightly different temperature evolution: from $\simeq13000\,\rm{K}$ to $\simeq7900\,\rm{K}$ in the first $\sim5\,\rm{days}$ of the spectroscopic evolution.

\ha~is the most prominent line.
From $\sim$\/$+14\,\rm{d}$ a second component appears, most likely \he~at 6678\AA, while \ha~becomes more evident. 
In Fig.~\ref{fig:radius}b we report the evolution of the expansion velocities inferred from the position of the minima of the P-Cygni profile. 
We infer expansion velocities for \h, declining from $\sim\,16500$\kms~at $+1.70\,\rm{d}$ to $\sim\,12200$\kms~at $\sim\,+21\,\rm{days}$.

At $\sim\,+3.93\,\rm{days}$ the \he~line at 5876\AA~becomes visible, with an expansion velocity decreasing from $\sim\,16500$ to $\sim\,8800$\kms~in $\sim\,21\,\rm{days}$.
The presence of helium in the early spectra of SN~2016gkg is also confirmed by our NIR spectra (Fig.~\ref{fig:spec}b).
Our $\sim\,+4\,\rm{d}$ and $\sim\,11\,\rm{d}$ spectra both show a blue continuum, with prominent Paschen lines in emission.
The emission at $\sim 10750$\AA~is most likely a blend of $\rm{Pa}\gamma$ and \he~at 10830\AA; the bluest absorption in the P-Cygni profile is consistent with the \he~expansion velocity inferred from the optical spectra.
From the shallow \he~(20581\AA) feature visible at $\sim\,+11\rm{d}$ and $\sim\,+15\,\rm{d}$ we infer an expansion velocity of $\sim10000$\kms, in agreement with our optical spectra. 

To support the classification of SN~2016gkg as a Type IIb SN, we use the Gelato \citep{2008A&A...488..383H} comparison tool on our $\sim\,+19\,\rm{d}$ spectrum (see Fig.~\ref{fig:spec}c), finding good matches with Type IIb SNe (SNe~1993J, 2008aq and 2011dh).

A detailed analysis, including the results from the complete spectroscopic follow-up campaign will be presented in an forthcoming paper.

\subsection{Progenitor properties}\label{sec:rad}
Soon after the shock breakout, the envelope is heated by the shock, expands and cools down. The time scale of this cooling phase depends on the initial progenitor radius, density profile, opacity and composition. 

We obtain a rough estimate on the progenitor radius of SN~2016gkg using the formalism of \citet{Rabinak11} (see their eq.~12) based on the temperature evolution within $\sim+5\,\rm{days}$ and a typical optical opacity for H-rich material \citep[$\kappa=0.34\,\rm{cm}^{2}\,\rm{g}^{-1}$][]{Rabinak11}. 
Assuming an explosion energy of $10^{51}\,\rm{erg}$ and a mass range of $\sim9-20$\msun (based on our {\it HST} data analysis; see Section~\ref{sec:prog} and Fig.~\ref{fig:prog2}), the progenitor of SN~2016gkg is consistent with a $\sim48-124$\rsun~star (Figure~\ref{fig:radius}a). 
This radius is consistent with \cite{Kilpatrick16}, who infer a radius of $\sim 70-650$\rsun~using the \cite{Rabinak11} formalism in conjunction with the early light curves of SN~2016gkg.
We did not consider the effects of a `dark phase' of the explosion \citep{2014ApJ...784...85P}, which would lead to an earlier explosion epoch and hence imply a larger radius for the progenitor of SN~2016gkg. \\
\begin{figure*}
\begin{center}
\mbox{ \epsfysize=8.0cm \epsfbox{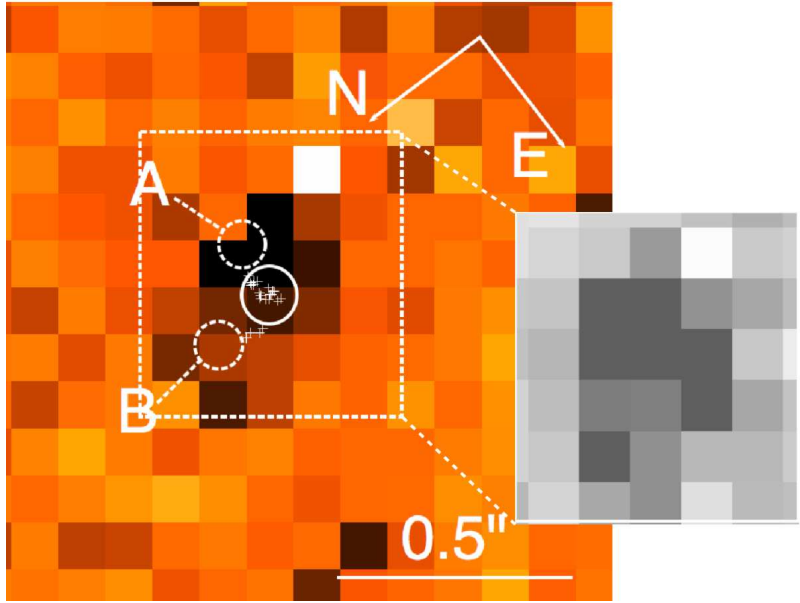}} 
\caption{The region around SN~2016gkg on the pre-explosion $F814W$ image and a zoom-in of the region with a different contrast, more clearly showing Source B. The white pixel is a masked bad pixel. Dashed circles indicate Sources A and B: the 0.05\arcsec\ radius corresponds to their positional uncertainty. The solid ellipse is the average of the transformed, jackknife-sampled positions (white crosses) of the SN, with the radius corresponding to the $1\sigma$ uncertainty (see text for details).
\label{fig:prog}}
\end{center}
\end{figure*}

\section{The progenitor star}\label{sec:prog}
We obtained adaptive optics imaging of SN~2016gkg with the Very Large Telescope (VLT) + Nasmyth Adaptive Optics System Near-Infrared Imager and Spectrograph (NaCo) on 2016 October 2.3 UT.
{\it Ks} imaging of $4500\,\rm{s}$ ($75\times60\,\rm{s}$) was taken with the S54 camera, using SN~2016gkg itself as a natural guide star.
Sky-subtracted images were aligned and co-added to produce a single deep post-explosion image.

Pre- and post-explosion {\it HST} and NaCo images were aligned using 7 common reference sources. 
All reference sources lie on one side of the SN position, leading to some extrapolation of the geometric transformation.
A transformation between the two set of pixel coordinates was derived allowing for translation, rotation and a single magnification factor.
Since so few sources were used for the alignment, we are susceptible to individual influential data points. 
Hence, we used a jack-knife sampling technique to test the effects of excluding either one or two reference sources from the fit. 
We used the derived transformation in each case to determine a position for the SN on the pre-explosion {\it F814W} image. The resulting position lies between two sources: Source A, at pixel 586.4, 789.4 and Source B, located at 585.9, 787.3 (see Fig.~\ref{fig:prog}).

The positions of A and B were measured using the {\sc dolphot} package, while their uncertainties ($\sim$0.5 WFPC2 pixels) were determined through Monte-Carlo simulations.
Source A is consistent with the $1\sigma$ error on the transformed SN position (including the r.m.s. errors from the geometric transformations, and the uncertainties estimated from jack-knife sampling), while B lies $\sim1.5\sigma$ distant. Both sources are hence viable progenitor candidates for SN~2016gkg. 

Both {\sc hstphot} and {\sc dolphot} \citep{Dolphin00} were used for the photometry. 
{\sc hstphot} only detects Source A, with magnitudes in the VEGAMAG system $F450W=23.75\pm0.19$, $F606W=23.56\pm0.10$, and $F814W=23.17\pm0.11\,\rm{mag}$. 
The source is well-fitted with a PSF, but has a sharpness parameter ($-0.31$) which is barely consistent with point sources (between $-0.3$ and $+0.3$), possibly implying that Source A is slightly extended.
{\sc dolphot} was run with standard parameters on both images in each filter simultaneously, and detects both sources. 
We find $F450W=23.60\pm0.14$, $F606W=23.72\pm0.08$, and $F814W=23.25\pm0.14\,\rm{mag}$ for A and $F450W=24.52\pm0.37$, $F606W=24.57\pm0.15$ and $F814W=24.13\pm0.29\,\rm{mag}$ for B.
Sharpness and $\chi^2$ are consistent with a single point source for both.
For Source A, we adopt the magnitudes from HSTPhot, as this is a fully optimized tool specifically developed for WFPC2.
We find different values for the progenitor photometry to \cite{Kilpatrick16} (after converting their photometry from STMAG to VEGAMAG: $F405W=23.42$, $F606W=23.10$, $F814W=23.32\,\rm{mag}$), and so we performed additional checks. The progenitor is in the Hubble Source Catalog\footnote{\url{https://archive.stsci.edu/hst/hsc/}} with magnitude $F450W=23.85\pm0.08\,\rm{mag}$ (converted to VEGAMAG from ABmag) and $F606W=23.34\pm0.05\,\rm{mag}$ which is marginally brighter than the $23.56\pm0.10\,\rm{mag}$ we find. We also compared the $F606W$ magnitudes of a set of sources on the WF4 chip in the HSC and from our own measurements with HSTPhot, and found no systematic offset between the two. 

\begin{figure*}
\begin{center}
\mbox{ \epsfysize=7.0cm \epsfbox{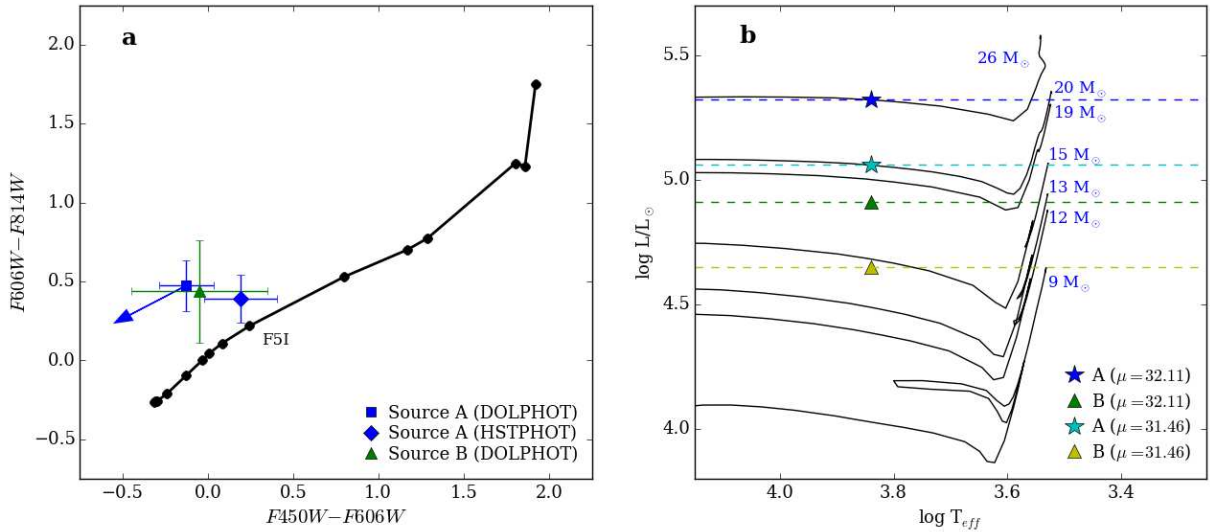}} 
\caption{{\bf a:} Color-color diagram showing the positions of sources A and B with respect to the theoretical locus of SG stars obtained using ATLAS9 models with solar metallicities. A reddening vector corresponding to an extinction of $A_V=1\,\rm{mag}$ is also shown. {\bf b:} Hertzsprung-Russell diagram showing the positions of A and B with respect to evolutionary tracks computed using the \textsc{stars} models, assuming solar metallicity.
We also show the effects of the large uncertainties on the distance modulus (see text for details).
\label{fig:prog2}}
\end{center}
\end{figure*}

In Fig.~\ref{fig:prog2}a we show the positions of A and B in a color-color diagram, with respect to the theoretical locus of supergiants (SG) derived using {\sc synphot} from ATLAS9 models \citep{2004astro.ph..5087C}: while their colors are slightly offset from the models, they are broadly consistent with a yellow or blue supergiant. \\

We did not correct the magnitudes for foreground Galactic ($A_V=0.05\,\rm{mag}$) or host extinction, since it is small compared to the uncertainty on distance, and will have minimal effect on the bolometric correction (see also Fig.~\ref{fig:prog2}a). In addition, we cannot exclude the presence of significant circumstellar dust around the progenitor, which would then be destroyed during the shock breakout \citep[e.g.][]{2012ApJ...759L..13F}. 

The large uncertainty on the distance of NGC~613 plays an important role in the identification of a plausible progenitor for SN~2016gkg.
A variety of distances are in the literature, spanning from $\sim$19.6 to $\sim31.7\,\rm{Mpc}$ ($\mu=31.46\pm0.80$ and $\mu=32.51\pm0.47\,\rm{mag}$ respectively).
In Fig.~\ref{fig:prog2}b, we show the position of A in the Hertzsprung-Russell diagram (HRD) with respect to a set of evolutionary tracks computed with the Cambridge \textsc{stars} models \citep{2004MNRAS.353...87E}. Adopting a distance modulus $\mu=32.11\pm0.38\,\rm{mag}$, a solar metallicity, the recommended values of temperature and surface gravity for F5I stars \citep[$T_{\rm{eff}}=6900\,\rm{K}$ and $\log{g}=+1.44\,\rm{dex}$][]{LandoltBornstein1982} and assuming a $0\,\rm{mag}$ bolometric correction \citep[see e.g.][]{2008PASP..120..583G}, we obtain $M_{bol}=-8.55\pm0.40,\rm{mag}$. 
Although A and B sit close to the 26\msun~and 15\msun~tracks respectively, none of them is a plausible SN progenitor, since they correspond to stars during the He core burning phase \citep[see][]{2009MNRAS.395.1409S}. 
Alternatively, the yellow colour could be explained by additional stripping experienced by the progenitor from a stellar companion \citep[see also][]{Kilpatrick16}.

Hence, we compare their luminosities with the endpoint of the evolutionary tracks (dashed lines in Fig.~\ref{fig:prog2}b).
Source A has $\log{L/L_{\sun}}=5.32\,\rm{dex}$, corresponding to the $19-20$\msun~tracks, while B falls in the region where the $12-13$\msun~tracks end, with $\log{L/L_{\sun}}=4.91\,\rm{dex}$.
These luminosities and temperatures estimates give radii of $\sim320$\rsun~and $200$\rsun~for A and B respectively.
If A is the progenitor star of SN~2016gkg, this would imply a yellow SG (YSG) more luminous than the progenitor of SN~1987A \citep{1987Natur.327...36W,1989A&A...219..229W}.
On the other hand, adopting $\mu=31.46\pm0.80$ ($D=19.6\,\rm{Mpc}$) for NGC~613, we obtain $M_{bol,A}=-7.90\pm0.81\,\rm{mag}$ and $M_{bol,B}=-6.89\pm0.81\,\rm{mag}$, corresponding to $\log{L_A/L_{\sun}}\simeq5.06\,\rm{dex}$ and $\log{L_B/L_{\sun}}\simeq4.65\,\rm{dex}$.
This would imply lower masses and radii for both Source A and B ($\sim15$\msun, with $\rm{R}\simeq240$\rsun~and $\sim9$\msun, with $\rm{R}\simeq150$\rsun~respectively).
This analysis suggests a larger radius for the progenitor of SN~2016gkg, arguing against the predictions of the \cite{Rabinak11} model.

Given the large uncertainties, we have no arguments to favor a particular distance for NGC~613, and we cannot exclude any of the mentioned values of the mass for the candidate progenitor.
Also, given the SN position in the {\it HST} images, we lack the astrometric accuracy to exclude either of the two sources as a progenitor for SN~2016gkg, rule out the presence of a stellar companion \citep[see][]{Kilpatrick16} or even exclude the possibility that the sources we have discussed are in fact star clusters. 
An improved distance to NGC 613 will address the first issue, while late time imaging will clarify direct progenitor constraints. 

\section{Summary \& Conclusions}  
The spectra of SN~2016gkg closely resemble those of other Type IIb SNe, in particular SNe~1993J and 2011dh.
\he~lines, the classical features for this class of transients, are visible since $+8.82\,\rm{d}$, becoming stronger with time.

Our photometric analysis reveals double peaked light curves in all optical and UV bands, with a relatively fast decline in $g-$ and $B-$ bands, faster than in SN~1993J, but comparable to SN~2011dh.

Deep {\it HST} pre-explosion images, and the temperature evolution within the first $\sim5\,\rm{days}$ from explosion, help to constrain the physical properties of the progenitor.
Following \cite{Rabinak11} we obtain a radius consistent with a $\sim48-124$\rsun~supergiant.
The analysis of the archival {\it HST} images revealed the presence of two sources at the position of SN~2016gkg (Source A and B).
Source A seems to correspond to a yellow supergiant star with an initial mass and radius in the range $15-20$\msun~ and $240-320$\rsun, while B corresponds to a $9-13$\msun~star with $R\simeq150-200$\rsun.
These radii are larger than those obtained adopting the \cite{Rabinak11} formalism.
Like in SN~2011dh, this might be due to the progenitor structure required to produce the double peaked light curve observed in Type IIb SNe \citep[i.e. the presence of an extended, $R\simeq10^{13}\,\rm{cm}$, low mass, $M\leq0.1$\msun~envelope becoming transparent after a few days of expansion, see][]{Nakar14,2012ApJ...757...31B}.

Given the large uncertainties on the distance of NGC~613, and the unfavorable position in the {\it HST} images, we cannot rule out either of the two sources as the possible progenitor star, as well as the presence of a stellar companion.
Deep late-time post-explosion {\it HST} images centered on the position of SN~2016gkg will provide the final word on the nature of the progenitor system.

\acknowledgments
\noindent Based on data from: 
ESO as part of PESSTO (197.D-1075.191.D-0935.) and under programme 097.D-0762(A).
Gemini Observatory under GS-2016B-Q-22 (PI: Sand). 
The Nordic Optical Telescope of the Instituto de Astrof\'isica de Canarias. 
The European Organisation for Astronomical Research in the Southern Hemisphere under ESO programme 198.A-0915(A). \\
FOSCGUI is a graphic user interface aimed at extracting SN spectroscopy and photometry obtained with FOSC-like instruments, developed by E. Cappellaro. \\
D.J.S. acknowledge NSF grants AST-1412504 and AST-1517649.
S.J.S. acknowledge EU/FP7-ERC grant 29122.
M.F. acknowledges the support of a Royal Society -- Science Foundation Ireland University Research Fellowship.
L.G. was supported in part by the US National Science Foundation under Grant AST-1311862.
M.S. acknowledges support from EU/FP7-ERC grant number 615929.
K.M. acknowledges support from STFC through an Ernest Rutherford Fellowship.
D.A.H., C.M., and G.H. are funded by NSF AST-1313484.
Support for IA was provided by NASA through the Einstein Fellowship Program, grant PF6-170148.
N.E.R. acknowledges financial support by the 1994 PRIN-INAF 2014 (project ``Transient Universe: unveiling new types of stellar explosions with PESSTO") and by MIUR PRIN 2010- 2011, ``The dark Universe and the cosmic evolution of baryons: from current surveys to Euclid".
M. S. acknowledges support from the Danish Agency for Science and Technology and Innovation via a Sapere Aude Level 2 grant, the Villum foundation, and the Instrumentcenter for Danish Astrophysics (IDA).

\end{document}